\documentclass[11pt]{article}
\usepackage{amsmath, amssymb,graphicx}

\newsavebox{\PSLASH}
\sbox{\PSLASH}{$p$\hspace{-1.8mm}/}

\begin{document}
\title{ Area Distribution of Elastic Brownian Motion  }
\author{M. A. Rajabpour$^{a}$\footnote{e-mail: Rajabpour@to.infn.it} \\ \\
  $^{a}$Dip. di Fisica Teorica and INFN,
Universit{\`a} di Torino, Via P. Giuria 1, 10125 Torino,
\\Italy}
\maketitle
\begin{abstract}
We calculate the excursion and meander area distributions of the
elastic Brownian motion  by using the self adjoint extension of the
Hamiltonian of the free quantum particle on the half line. We also
give some comments on the area of the Brownian motion bridge on the
real line with the origin removed. We will stress on the power of
self adjoint extension to investigate different possible boundary
conditions for the stochastic processes. We discuss also some
possible physical applications.\vspace{5mm}

\noindent PACS number(s): 02.50.FZ and 03.65 Db and 05.40.Jc
\end{abstract}
\section{Introduction}\
\setcounter{equation}{0}\

In this paper we study some area distributions of  elastic Brownian
motion \cite{RY}, such  as elastic Brownian excursion and elastic
Brownian meander. The goal apart from the calculation of the area
distribution of elastic Brownian motion is to give a unified
framework to study different area distributions for Brownian motion
with generic boundary conditions. This generalization is in close
connection with the concept of the self adjoint extension in quantum
mechanics. The self adjoint extension gives a reasonable
classification for the possible boundary conditions of the quantum
particle and so for the stochastic process.

Different area distributions of Brownian motion in one dimension
were calculated by the mathematicians in the last century. The
Brownian motion area was first calculated by Kac \cite{Kac}. The
Brownian excursion area was calculated by Darling \cite{Darling} and
Louchard \cite{Louchard} and  the Brownian meander area was
calculated by Perman and Wellner \cite{perman}. For the applications
and other distributions of the areas see \cite{Flajolet1,Flajolet2}
and also the complete review by Janson \cite{Janson}, and references
therein.

The quantum mechanics methods first used by Comtet and Majumdar
\cite{CM} to re-drive the different area distributions of Brownian
motion. This method which is more common in the physics literature
apart from simplicity can give a unique way to calculate the
different area distributions. It is also very useful to generalize
the results to more complicated stochastic processes. The self
adjoint extension of the Hamiltonian operator gives us the variety
of different possible boundary conditions in the presence of the
boundary \cite{Rajabpour}. In this letter we will focus on the area
distribution of the elastic Brownian motion which is equivalent to
the free quantum particle on the half line. We also calculate the
area distribution of the stochastic process equivalent to the free
particle on the real line with a hole at the origin. Of course the
area distribution is just one of the myriad of possible
distributions that one can investigate for the elastic Brownian
motion and the Brownian motion on the pointed real line. We will
summarize some of these distributions in the last section.

The paper is organized as follows: in the second section we define
the elastic Brownian motion and its connection to the quantum
particle on the half line. In the third section we use the same
method as \cite{CM} to calculate the area distribution of the
elastic Brownian excursion and meander. Different limits of our
calculation will give the well known results. We will conclude this
section with some immediate application of our results and giving
some hints about other useful Brownian functionals. In the forth
section using the results of the section three we will give the
distribution of the area for Brownian bridge with a point defect at
the origin. Finally in the last section we will summarize our
results and possible future directions.

\section{Elastic Brownian motion and quantum mechanics}\
\setcounter{equation}{0}\

The elastic Brownian motion is the natural generalization of the
Brownian motion in the half line with special boundary condition on
the origin. To define the process we need first to introduce the
local time. The definition of the local time of the path $\omega$ at
the point $a$ is as follows
\begin{eqnarray}\label{local time}
t_{l}(a):=\frac{1}{2}\lim_{\epsilon\rightarrow0}\frac{1}{\varepsilon}\int_{0}^{T}\textbf{1}_{x+\varepsilon}(B_{s})ds,
\end{eqnarray}
where $\textbf{1}_{x+\varepsilon}(B_{s})$ is the indicator for the
time that the process is in the interval $[0,x+\epsilon]$. One can
naively write the above equation as an integral over a delta
function as $t_{l}(\omega,0)=\int_{0}^{T}\delta(x(t))dt$. The local
time has dimensions of the inverse of velocity and can be written as
$t_{l}(a)=n\frac{\Delta t}{\Delta x}$ for the discrete random walk,
where $n$ is the number of times the path hits the origin. Then one
can write the exponential of the local time as
\begin{eqnarray}\label{exponential of local time}
\exp{(\frac{-2\pi}{\eta}t_{l})}\approx (1-\frac{2\pi}{\eta}\Delta
x)^{n}.
\end{eqnarray}
To go to the discrete level we multiplied $t_{l}$ with
$\frac{(\Delta x)^{2}}{\Delta t}$ that comes from the central limit
theorem. The equation (\ref{exponential of local time}) means that
by considering $1-\frac{2\pi}{\eta}t_{l}$ as the probability for a
single reflection from the origin it is possible to interpret the
exponential of local time as the probability that a particle on a
given path is reflected from the origin\footnote{The name elastic
Brownian motion comes from this property of the process.
$\eta=\infty$ is the reflecting barrier, the particle will be
reflected from the boundary with probability one. $\eta=0$ is the
absorbing barrier, the particle will be absorbed after hitting the
boundary. Other cases between two extreme cases called elastic
barrier.}. Then the Green function of the elastic Brownian motion is
just by the following expectation value
\begin{eqnarray}\label{elastic bm}
G(x,y,T)=<\exp{(\frac{-2\pi}{\eta}t_{l})}>.
\end{eqnarray}
The elastic Brownian motion is in close connection with the quantum
particle on the half line. The Hamiltonian operator of the quantum
particle on the half line has self adjoint extension with the
following boundary condition
\begin{eqnarray}\label{boundary condition for d=3}
\psi(0)=\frac{-\eta}{2\pi}\frac{d\psi}{dx}|_{x=0}.
\end{eqnarray}
The above boundary condition is called Robin boundary condition. The
energy of the particle is $E_{k}=\frac{1}{2}k^{2}$ and the wave
functions are
\begin{eqnarray}\label{Wave function momentu}
\psi_{k}=\sqrt{\frac{2}{\pi}}\cos(kx+\delta_{k}),
\end{eqnarray}
where $\tan(\delta_{k})=\frac{2\pi}{\eta k}$ and $\delta_{k}$ is the
phase shift corresponding to the $s$-wave. The solutions are
normalizable and complete.

The Hamiltonian is self adjoint for all of the real values of $\eta$
but to avoid the cases with bound states we will consider just
non-positive $\eta$.  $\eta=0$ is the Dirichlet boundary condition
and $\eta\rightarrow-\infty$ is the Neumann boundary condition. The
Green function with respect to the solutions of the Hamiltonian has
the following form
\begin{eqnarray}\label{green function}
G_{\eta}(x,y,t)=\int_{0}^{\infty}dke^{-iE_{k}t}\varphi(y)\varphi^{*}(x).
\end{eqnarray}
Using the above equation the Green function has the following form
for arbitrary values of the self adjoint extension
\cite{clark.menikoff.sharp,gutman1,gutman2}
\begin{eqnarray}\label{green function in three}
G_{\eta}(x,y,t)&=&G_{F}(x-y,t)+G_{F}(x+y,t)+\frac{4\pi}{\eta}\int_{0}^{\infty}dwe^{\frac{2\pi}{\eta}w}G_{F}(x+y+w)\hspace{0.3cm}\eta\leq0;\qquad\\
G_{\eta}(x,y,t)&=&G_{F}(x-y,t)+G_{F}(x+y,t)-\frac{4\pi}{\eta}\int_{0}^{\infty}dwe^{\frac{-2\pi}{\eta}w}G_{F}(x+y-w)\nonumber\\
&+&\frac{4\pi}{\eta}e^{i\frac{2\pi^{2}t}{\eta^{2}}}e^{-\frac{2\pi}{\eta}(x+y)}\hspace{1cm}\eta\geq0;\\
G_{F}(x-y,t)&=&\frac{1}{\sqrt{2\pi it}}e^{i(x-y)^{2}/2t}.
\end{eqnarray}
For the special cases, Dirichlet and Neumann the results are as
follows
\begin{eqnarray}\label{Dirichlet Newman}
G_{\eta=0}(x,y,t)=G_{F}(x-y,t)-G_{F}(x+y,t);\\
G_{\eta\rightarrow-\infty}(x,y,t)=G_{F}(x-y,t)+G_{F}(x+y,t).
\end{eqnarray}
One can use the above equations to get the Green function of the
elastic Brownian motion by just Wick rotation. The important point
of this section is the possibility of using quantum mechanics
language to describe the elastic Brownian motion. The other
interesting point is the possibility of extending this equality in
to the level of path integral representation
\cite{clark.menikoff.sharp,gutman1,gutman2}. In the next section we
use this correspondence to calculate different area distributions of
the elastic Brownian motion.

\section{Area of the elastic Brownian motion }\

In this section we will solve the problem of the area for the
restricted Brownian motion, in particular we will solve the problem
of the area distribution of elastic Brownian excursion and elastic
Brownian meander. In the extreme limits we will get the well known
results.
\subsection{The Area Under the elastic Brownian excursion}
\setcounter{equation}{0}\
 In this subsection we calculate
excursion area of the elastic Brownian motion. The definition of the
excursion area is as follows: take an elastic Brownian process
starts at $x(0)=\epsilon$ and comes back after time $T$ to
$x(T)=\epsilon$, without crossing the origin between. We are
interested to the probability density $P(A,T,\epsilon)$ of the area
$A=\int_{0}^{T}x(\tau)d\tau$ for a fixed $\epsilon$ and then finally
take the limit $\epsilon\rightarrow 0$, it plays the role of the
regulator and can be treated independent of the limit in the local
time process. This regularization is just necessary for the
Dirichlet boundary condition but we are happy to keep it in our
calculation to see its relevance in the calculation of Dirichlet
boundary condition. To do this calculation we map the problem to the
quantum mechanical problem. Now we use the method of Comtet and
Majumdar \cite{CM} to calculate the excursion area distribution. To
satisfy the constraint that process stays non-negative between $0$
and $T$ one can multiply the above measure with the indicator
function
$\prod_{\tau=0}^{T}\theta[x(\tau)]\exp{(\frac{-2\pi}{\eta}t_{l})}$
which $\theta$ is the step function. The distribution $P(A,T)$ of
the area under the elastic Brownian excursion can then be expressed
as the following quantum mechanical problem
\begin{eqnarray}\label{path integral for PA}
P^{\eta}(A,T)=\frac{1}{Z^{\eta}_{E}}\int_{x(0)=\epsilon}^{x(\tau)=\epsilon}\mathcal{D}x(\tau)
e^{-\int_{0}^{T}(\frac{1}{2}(\frac{dx}{d\tau})^{2}+\frac{2\pi}{\eta}\delta(x(t))}\prod_{\tau=0}^{T}\theta[x(\tau)]\delta(\int_{0}^{T}x(\tau)d\tau-A),
\end{eqnarray}
where $Z^{\eta}_{E}$ is the normalization and corresponds to the
quantum mechanics of a particle with infinite wall at the origin and
zero potential at the positive real line
\begin{eqnarray}\label{ZE1}
Z^{\eta}_{E}=<\epsilon\mid e^{-H_{0}T}\mid \epsilon>.
\end{eqnarray}
The Hamiltonian, $H_{0}$ is equal to the self adjoint Hamiltonian
that we discussed in the previous section. After integration, for
small $\epsilon$ we have
\begin{eqnarray}\label{ZE2}
Z^{\eta}_{E}\simeq 2(\eta-2\pi\epsilon)^{2}(\frac{1}{\sqrt{2\pi
T}\eta^{2}}-\frac{\pi}{\eta^{3}}e^{\frac{2\pi^{2}T}{\eta^{2}}}\mbox{Erfc}(\sqrt{2T}\frac{\pi}{\eta})).
\end{eqnarray}
The integral for the Dirichlet and Neumann cases are
\begin{eqnarray}\label{DN}
Z^{0}_{E}&\simeq& \frac{1}{\sqrt{2\pi}}\epsilon^{2}T^{-3/2}+\textit{O}(\epsilon^{3});\hspace{1cm} \eta=0,\nonumber\\
Z^{\infty}_{E}&\simeq& \sqrt{\frac{2}{\pi
T}}+\textit{O}(\epsilon^{2});\hspace{2cm}\eta\rightarrow-\infty.
\end{eqnarray}
The above partition functions are just the probability that an
elastic Brownian motion goes from $x(0)=\epsilon$ to
$x(\tau)=\epsilon$ in time $T$ without crossing the origin. Taking
the Laplace transform
$P(\lambda,T)=\int_{0}^{\infty}P(A,T)e^{-\lambda A}dA$ of both sides
of the the equation (\ref{path integral for PA}) gives
\begin{eqnarray}\label{path integral for PA laplace space}
P^{\eta}(\lambda,T)=\frac{1}{Z_{E}}\int_{x(0)=\epsilon}^{x(\tau)=\epsilon}\mathcal{D}x(\tau)
e^{-\int_{0}^{T}(\frac{1}{2}(\frac{dx}{d\tau})^{2}+\frac{2\pi}{\eta}\delta(x(t))+\lambda
x(\tau))}\prod_{\tau=0}^{T}\theta[x(\tau)].
\end{eqnarray}
In the numerator we have the propagator $<\epsilon\mid
e^{-H_{1}T}\mid\epsilon>$ where
$H_{1}=-\frac{1}{2}(\frac{dx}{d\tau})^{2}+V(x)$ and $V(x)=\lambda x$
for $x>0$ and infinite for $x\leq0$. We absorb the Dirac delta
function in to the boundary condition as the case of quantum
particle on the half line, in the other words we consider the self
adjoint extension of this operator. The boundary condition of the
wave function after self adjoint extension is the same as
(\ref{boundary condition for d=3}). The solution of the
Schr\"{o}dinger equation is given by the Airy function as follows
\begin{eqnarray}\label{Airy wave function}
\psi^{\eta}_{i}(x)=\frac{Ai((2\lambda)^{1/3}(x-E/\lambda))}{\sqrt{\int_{0}^{\infty}Ai^{2}((2\lambda)^{1/3}(y-E/\lambda))dy}}.
\end{eqnarray}
Using the boundary conditions one can determine the discrete
eigenvalues as follows
\begin{eqnarray}\label{Airy eigenvalues}
E^{\eta}_{i}=2^{-1/3}\lambda^{2/3}c^{\eta}_{i},\hspace{1cm}\frac{Ai(-c^{\eta}_{i})}{Ai'(-c^{\eta}_{i})}=-\frac{\eta
(2\lambda)^{1/3}}{2\pi}.
\end{eqnarray}
Unfortunately since the above equation is transcendental the exact
form of $c^{\eta}_{i}$ for the generic boundary condition is not
available. However, for the Dirichlet and Neuman boundary conditions
the solutions are just the magnitude of the zeros of $Ai(z)$ and
$Ai'(z)$ on the negative real axes respectively, we show them by
$-c^{0}_{i}$ and $-c^{\infty}_{i}$. The first few real roots of
$Ai(z)$ are approximately ~-2.33811, ~-4.08795, ~-5.52056 and
~-6.78671, etc. The first few real roots of $Ai'(z)$ are
approximately ~-1.018792, ~- 3.248197, ~-4.820099 and ~-6.163307,
etc. Then the wave functions in these two cases are
\begin{eqnarray}\label{Airy wave function DN}
\psi^{0}_{i}(x)=(2\lambda)^{1/6}\frac{Ai((2\lambda)^{1/3}x-c^{0}_{i})}{Ai'(-c^{0}_{i})},\\
\psi^{\infty}_{i}(x)=(2\lambda)^{1/6}\frac{Ai((2\lambda)^{1/3}x-c^{\infty}_{i})}{\sqrt{c^{\infty}_{i}}Ai(-c^{\infty}_{i})}.
\end{eqnarray}
To get the above results we used the identity
$\int_{x}^{\infty}Ai^{2}(x)dx=-xAi^{2}(x)+Ai'^{2}(x)$. Using the
energy eigenvalues and wave functions one can write the equation
(\ref{path integral for PA laplace space}) as follows

\begin{eqnarray}\label{PA laplace space}
P^{\eta}(\lambda,T)=\frac{<\epsilon\mid e^{-H_{1}T}\mid
\epsilon>}{Z^{\eta}_{E}}=\frac{1}{Z^{\eta}_{E}}\sum_{i=1}^{\infty}|\psi(\epsilon)|^{2}e^{-2^{-1/3}\lambda^{2/3}c^{\eta}_{i}T}.
\end{eqnarray}
For the Dirichlet and Neumann cases after considering small
$\epsilon$ we have
\begin{eqnarray}\label{PA laplace space final}
P^{0}(\lambda,T)&=&\sqrt{2\pi}\lambda
T^{3/2}\sum_{i=1}^{\infty}e^{-2^{-1/3}\lambda^{2/3}c^{0}_{i}T},\\
P^{\infty}(\lambda,T)&=&\sqrt{\frac{\pi T}{2}}(2\lambda)^{1/3}
\sum_{i=1}^{\infty}\frac{1}{c^{\infty}_{i}}e^{-2^{-1/3}\lambda^{2/3}c^{\infty}_{i}T}.
\end{eqnarray}
One can write the equations (\ref{PA laplace space final}) and $(3.12)$ in the
scaling form as follows
\begin{eqnarray}\label{PA laplace space final scaling form}
P^{0}(\lambda,T)&=&s\sqrt{2\pi}\sum_{i=1}^{\infty}e^{-2^{-1/3}s^{2/3}c^{0}_{i}},\\
P^{\infty}(\lambda,T)&=&u\sqrt{\frac{\pi }{2}}2^{1/3}
\sum_{i=1}^{\infty}\frac{1}{c^{\infty}_{i}}e^{-2^{-1/3}u^{2}c^{\infty}_{i}},
\end{eqnarray}
where $s=\lambda T^{3/2}$ and $u=T^{1/2}\lambda^{1/3}$. It is
possible to do the inverse Laplace transform of the functions
(\ref{PA laplace space final}) and $(3.12)$ explicitly and find the distribution
of the excursion area. To do so we need the formula
\begin{eqnarray}\label{Laplace transform1}
\mathcal{L}^{-1}[\exp(-s\lambda^{a});A]=\frac{as}{A^{a+1}}M_{a}(sA^{-a})
\end{eqnarray}
where $M_{\frac{\beta}{2}}(z)$ is the well known Wright function
given by the following series
\begin{eqnarray}\label{wright1}
M_{\frac{\beta}{2}}(z)=\frac{1}{\pi}\sum_{k=0}^{\infty}\frac{(-z)^{k}}{k!}\Gamma(\beta\frac{(k+1)}{2})\sin(\beta\frac{(k+1)}{2}).
\end{eqnarray}
For the Dirichlet boundary condition the inverse Laplace transform
for $T=1$ gives
\begin{eqnarray}\label{alfa2 pa}
P^{0}(A)=\sqrt{2\pi}\frac{2^{\frac{2}{3}}}{3}\sum_{k=1}^{\infty}c^{0}_{i}
\partial_{A}(A^{-\frac{5}{3}}M(\frac{c^{0}_{i}}{2^{\frac{1}{3}}}A^{-\frac{2}{3}},\frac{2}{3}))=\nonumber\\
\frac{2\sqrt{6}}{A^{\frac{10}{3}}}\sum_{k=1}^{\infty}e^{-\frac{2(c^{0}_{i})^{3}}{27A^{2}}}(\frac{2(c^{0}_{i})^{3}}{27})^{2/3}U(-\frac{5}{6},\frac{4}{3},\frac{2(c^{0}_{i})^{3}}{27A^{2}}).
\end{eqnarray}
where $U(a,b,z)$ is the confluent hypergeometric function.

To calculate the moments of the area one can  work in the Laplace
space and then use the equality
$\Gamma(n)<A^{-n}>=\int_{0}^{\infty}P(\lambda,T)\lambda^{n-1}d\lambda$.
 To do the calculations we need  to first define the generalized
Riemann function
$\Lambda^{\eta}(s)=\sum_{i=1}^{\infty}\frac{1}{(c^{\eta}_{i})^{s}}$
where $c^{\eta}_{i}$ comes from the equation (\ref{Airy
eigenvalues}). The above relations help us to calculate the moments
explicitly as follows
\begin{eqnarray}\label{moments Brownian1}
<A^{n}>=\sqrt{2\pi}2^{\frac{1-n}{2}}\frac{n\Gamma(1+\frac{3(1-n)}{2})}{\Gamma(2-n)}\Lambda^{0}(\frac{3(-n+1)}{2}).
\end{eqnarray}
It was shown in \cite{Takacas} that the moments after regularization
are
\begin{eqnarray}\label{moments Brownian2}
<A^{n}>=\sqrt{2\pi}2^{\frac{4-n}{2}}\frac{\Gamma(n+1)}{\Gamma(\frac{3n-1}{2})}K_{n}
\end{eqnarray}
where $K_{n}$ is by the following recursion relation

\begin{eqnarray}\label{K series}
K_{n}=\frac{3n-4}{4}K_{n-1}+\sum_{j=1}^{n-1}K_{j}K_{n-j},
\hspace{1cm}s\geq1,
\end{eqnarray}
the first few values are $K_{0}=-\frac{1}{2}$, $K_{1}=\frac{1}{8}$,
$K_{2}=\frac{5}{64}$ and $K_{3}=\frac{15}{128}$. Then the first few
moments of the excursion area are
\begin{eqnarray}\label{Area moments of excursion}
<A^{0}>=1,
\hspace{0.5cm}<A^{1}>=\frac{\sqrt{2\pi}}{4},\hspace{0.5cm}<A^{2}>=\frac{5}{12},\hspace{0.5cm}
<A^{3}>=\frac{15\sqrt{2\pi}}{128},....
\end{eqnarray}
There is also a nice relation between the Airy zeta function and
$K_{n}$ as follows
\begin{eqnarray}\label{K and lamda}
\Lambda^{0}(\frac{3-3n}{2})=-\frac{4}{3}\frac{\cos(\frac{3\pi
n}{2})}{\sin(\pi n)}K(n).
\end{eqnarray}
For example we have the following limits
\begin{eqnarray}\label{ lamda}
\lim_{n\rightarrow0}
n\Lambda^{0}(\frac{3}{2}(n-1))=\frac{2}{3\pi},\hspace{1cm}\Lambda^{0}(0)=\frac{1}{4}.
\end{eqnarray}
For the Neumann boundary condition we need another Laplace transform
pair
\begin{eqnarray}\label{Laplace pairs}
\mathcal{L}^{-1}[\lambda^{-\alpha}\exp(-s\lambda^{-a});A]=\frac{1}{A^{1-\alpha}}\phi(a,\alpha;-sA^{a});
\hspace{0.3cm}-1<a<0, \hspace{0.3cm}s>0, \hspace{0.3cm}0<\alpha<1,
\end{eqnarray}
where
$\phi(a,\alpha;z)=\sum_{k=0}^{\infty}\frac{z^{k}}{k!\Gamma(ak+\alpha)}$
is the generalized Wright function defined for the $a>-1$. Using the
above formula the area distribution has the following form
\begin{eqnarray}\label{Pa Neumann}
P^{\infty}(A)=\sqrt{\frac{\pi
T}{2}}2^{1/3}\sum_{i=1}^{\infty}\frac{1}{c^{\infty}_{i}}\frac{\partial}{\partial
A}(\frac{1}{A^{1/3}}\phi(\frac{-2}{3},\frac{2}{3};-2^{-1/3}c^{\infty}_{i}TA^{\frac{-2}{3}})).
\end{eqnarray}
The moments of the area for $T=1$ can be written as
\begin{eqnarray}\label{moment area N1}
<A^{n}>=\frac{3\sqrt{\pi}}{2^{\frac{n+1}{2}}}\frac{\Gamma(\frac{1-3n}{2})}{\Gamma(-n)}\Lambda^{\infty}(\frac{3-3n}{2}).
\end{eqnarray}
These moments are the same as the moments of the Brownian bridge and
can be regularized in the same way \cite{Janson}. Then the moments
are
\begin{eqnarray}\label{moment area N2}
<A^{n}>=\frac{\sqrt{\pi}2^{-n/2}\Gamma(1+n)}{\Gamma(\frac{1+3n}{2})}D_{n},
\hspace{1cm}n\geq0,
\end{eqnarray}
where $D_{n}$ is by the following recursion relation

\begin{eqnarray}\label{K series}
D_{n}=\frac{3n-2}{4}D_{n-1}-\frac{1}{2}\sum_{j=1}^{n-1}D_{j}D_{n-j},
\hspace{1cm}s\geq1.
\end{eqnarray}
The first few values are $D_{0}=1$, $D_{1}=\frac{1}{4}$,
$D_{2}=\frac{7}{32}$ and $D_{3}=\frac{21}{64}$. Then the first few
moments of the  area are
\begin{eqnarray}\label{Area moments of bridge}
<A^{0}>=1,
\hspace{0.5cm}<A^{1}>=\frac{1}{4}\sqrt{\frac{\pi}{2}},\hspace{0.5cm}<A^{2}>=\frac{7}{60},\hspace{0.5cm}
<A^{3}>=\frac{21}{512}\sqrt{\frac{\pi}{2}},....
\end{eqnarray}

It is not possible to find exact probability distribution for the
generic $\eta$ because for Robin boundary condition $c^{\eta}_{i}$
is not independent of $\lambda$ and since they are related by
non-algebraic relation it is not possible to get $c^{\eta}_{i}$ with
respect to $\lambda$ explicitly. However one can follow the
calculations in the level of Laplace space. The wave function has
the following form
\begin{eqnarray}\label{generic wave function}
\psi^{\eta}_{i}(x)=\frac{(2\lambda)^{1/6}Ai((2\lambda)^{1/3}x-c^{\eta}_{i})}{\sqrt{c^{\eta}_{i}Ai^{2}(-c^{\eta}_{i})+Ai'^{2}(-c^{\eta}_{i})}}.
\end{eqnarray}
The Laplace transform of the distribution of the area is
\begin{eqnarray}\label{PA laplace space2}
P^{\eta}(\lambda,T)=\frac{(2\lambda)^{1/3}}{Z_{E}(\epsilon=0)}\sum_{i=1}^{\infty}(\frac{\eta^{2}(2\lambda)^{2/3}}{4\pi^{2}})(\frac{1}{1+(\frac{\eta^{2}(2\lambda)^{2/3}}{4\pi^{2}})c^{\eta}_{i}})e^{-2^{-1/3}\lambda^{2/3}c^{\eta}_{i}T}.
\end{eqnarray}
To pursuit the calculation let  us consider small $\eta$s. One can
write $c^{\eta}_{i}=c^{0}_{i}+\delta c^{\eta}_{i}$ where $\delta
c^{\eta}_{i}$ is the small perturbation around the zeros of the Airy
function. After expansion of (\ref{Airy eigenvalues}) the
perturbation is
\begin{eqnarray}\label{perturbation of energy1}
\delta c^{\eta}_{i}\approx\frac{\eta (2\lambda)^{1/3}}{2\pi}.
\end{eqnarray}
For small $\eta$ one can also expand $Z_{E}$ as follows
\begin{eqnarray}\label{ZE small eta}
Z^{\eta}_{E}\approx
\frac{\eta^{2}}{2^{3/2}\pi^{5/2}}+\mathcal{O}(\eta^{4}).
\end{eqnarray}
Then the $P(\lambda,T)$ after expansion is
\begin{eqnarray}\label{PA laplace space3}
P^{\eta}(\lambda,T)\approx\sqrt{2\pi}\lambda
T^{\frac{3}{2}}\sum_{i=1}^{\infty}e^{-2^{-1/3}\lambda^{2/3}c^{0}_{i}T-\frac{\eta
T}{2\pi}\lambda}.
\end{eqnarray}
The inverse Laplace transform of the function after using the
equation (\ref{Laplace transform1}) is
\begin{eqnarray}\label{alfa2 pa2}
P^{\eta}(A)\approx\sqrt{2\pi}\frac{2^{2/3}}{3}\sum_{k=1}^{\infty}c^{0}_{i}
\partial_{A}\left((A-\frac{\eta }{2\pi})^{-\frac{5}{3}}M(\frac{c^{0}_{i}}{2^{\frac{1}{3}}}(A-\frac{\eta }{2\pi})^{-\frac{2}{3}},\frac{2}{3})\right).
\end{eqnarray}
The moments of the area can be find by the same method as before by
just replacing $A$ in the equation (\ref{moments Brownian2}) by
$A-\frac{\eta }{2\pi}$.

For the large $\eta$s the same calculation can be done. The
partition function of the elastic Brownian motion for large $\eta$
is
\begin{eqnarray}\label{large eta}
Z^{\eta}_{E}\approx \sqrt{\frac{2}{\pi
T}}-\frac{2\pi}{\eta}+\mathcal{O}(\frac{1}{\eta^{2}}).
\end{eqnarray}
The perturbation of $c^{\eta}_{i}$ after expansion of the equation
(\ref{Airy eigenvalues}) is
\begin{eqnarray}\label{perturbation of energy2}
\delta
c^{\eta}_{i}\approx\frac{-2\pi}{c^{\infty}_{i}\eta(2\lambda)^{1/3}}.
\end{eqnarray}
Unfortunately since the perturbation of the energy is dependent on
the energy level we are not able to find simple equation for the
distribution of the area in this case.

\subsection{The area under the elastic Brownian meander}

The definition of the elastic Brownian meander is similar to the
elastic Brownian excursion, the only difference is for the elastic
Brownian meander we do not need to force the process to come back to
the starting point. In this case the partition function is
\begin{eqnarray}\label{Z menader1}
Z^{\eta}_{M}=\int_{0}^{\infty}db<b| e^{-H_{0}T}|\epsilon>.
\end{eqnarray}
One can show that
\begin{eqnarray}\label{Z menader2}
<b| e^{-H_{0}T}|\epsilon>&=&\sqrt{\frac{1}{2\pi
T}}\left(e^{-\frac{1}{2}\frac{(\epsilon-b)^{2}}{T}}+e^{-\frac{1}{2}\frac{(\epsilon+b)^{2}}{T}}\right)+\nonumber\\
&\frac{2\pi}{\eta}&e^{\frac{2\pi^{2}}{\eta^{2}}T-\frac{2\pi(\epsilon+b)}{\eta}}
Erfc
[(\frac{2\pi^{2}T}{\eta^{2}})^{1/2}-\frac{\pi(\epsilon+b)}{\eta}(\frac{2T\pi^{2}}{\eta^{2}})^{-1/2}].
\end{eqnarray}
For small $\eta$ one can expand the error function up to the second
order
\begin{eqnarray}\label{Z menader3}
<b| e^{-H_{0}T}|\epsilon>&\approx&\frac{1}{\sqrt{2\pi
T}}\left(e^{-\frac{1}{2}\frac{(\epsilon-b)^{2}}{T}}+e^{-\frac{1}{2}\frac{(\epsilon+b)^{2}}{T}}\right)-
\frac{2}{\sqrt{2\pi
T}}\frac{e^{-\frac{1}{2}\frac{(\epsilon+b)^{2}}{T}}}{(1-\eta\frac{\epsilon+b}{2\pi
T })}.
\end{eqnarray}
Taking the first order correction with respect to the $\eta$ and
integrating over $b$ gives
\begin{eqnarray}\label{Z menader4}
Z^{\eta}_{M}\approx
Erf(\frac{\epsilon}{\sqrt{2T}})-\frac{\eta}{\pi\sqrt{2\pi
T}}e^{-\frac{\epsilon^{2}}{2T}}.
\end{eqnarray}
For $\eta\rightarrow\infty$ it is easy to get
\begin{eqnarray}\label{Z menader Neumann}
Z^{\infty}_{M}\approx 1.
\end{eqnarray}
Similar to the calculation that we did in the  elastic Brownian
excursion case one can write the Laplace transform of the
distribution of area as
\begin{eqnarray}\label{Laplace tansform of area meander}
P^{\eta}(\lambda,T)=\frac{1}{Z^{\eta}_{M}}\int_{0}^{\infty}db<b|
e^{-H_{1}T}|\epsilon>.
\end{eqnarray}
Using the wave function (\ref{generic wave function}) one can get
\begin{eqnarray}\label{PA laplace space3}
P^{\eta}(\lambda,T)=\frac{1}{Z^{\eta}_{M}}\sum_{i=1}^{\infty}\frac{Ai((2\lambda)^{1/3}\epsilon-c^{\eta}_{i})\int_{-c^{\eta}_{i}}^{\infty}Ai(y)dy}{c^{\eta}_{i}Ai^{2}(-c^{\eta}_{i})+Ai'^{2}(-c^{\eta}_{i})}e^{-2^{-1/3}\lambda^{2/3}c^{\eta}_{i}T}.
\end{eqnarray}
After expansion of the function with respect to the $\epsilon$ and
$\eta$ the first correction appears in the spectrum as follows
\begin{eqnarray}\label{PA laplace space4}
P^{\eta}(\lambda,T)=\sqrt{\pi}2^{-1/6}(\lambda
T^{3/2})^{1/3}\sum_{i=1}^{\infty}B(c^{0}_{i})e^{-2^{-1/3}\lambda^{2/3}c^{\eta}_{i}T}.
\end{eqnarray}
where
$B(c^{0}_{i})=\frac{\int_{-c^{0}_{i}}^{\infty}Ai(y)dy}{Ai'(-c^{0}_{i})}$
and $c^{\eta}_{i}=c^{0}_{i}+\delta c^{\eta}_{i}$. The distribution
of the area after inverse Laplace transform is
\begin{eqnarray}\label{PA  space4}
P^{\eta}(A,T)=\sqrt{\pi}2^{-1/6}
T^{1/2}\sum_{i=1}^{\infty}B(c^{0}_{i})\frac{\partial}{\partial
A}(\frac{1}{(A-\frac{\eta}{2\pi})^{1/3}}\phi(\frac{-2}{3},\frac{2}{3};-2^{-1/3}c^{\eta}_{i}T(A-\frac{\eta}{2\pi})^{\frac{-2}{3}})).
\end{eqnarray}

The continuation of calculation is now straightforward we just need
to use the well known results for the Brownian meander. The moments
of the area for Brownian meander, i.e. $\eta=0$, for $T=1$ is
\begin{eqnarray}\label{moments Brownian meander}
<A^{n}>=\sqrt{\pi}2^{-n/2}\frac{\Gamma(n+1)}{\Gamma(\frac{3n+1}{2})}Q_{n}.
\end{eqnarray}
$Q_{n}$ satisfies the following recursion relations
\begin{eqnarray}\label{recursion relations Q}
Q_{n}&=&\beta_{n}-\sum_{j=1}^{n}\alpha_{j}Q_{n-j},\nonumber\\
\beta_{n}&=&\alpha_{n}+\frac{3}{4}(2n-1) \beta_{n-1},\\
\alpha_{n}&=&\frac{6^{-2n}}{\Gamma(n+1)}\frac{\Gamma(3n+1/2)}{\Gamma(n+1/2)}.\nonumber
\end{eqnarray}
The first few values are $Q_{0}=1$, $Q_{1}=\frac{3}{4}$,
$Q_{2}=\frac{59}{32}$ and $Q_{3}=\frac{465}{64}$. Then the first few
values of the moments of the area are
\begin{eqnarray}\label{Area moments of Brownian meander}
<A^{0}>=1,
\hspace{0.5cm}<A^{1}>=\frac{3}{4}\sqrt{\frac{\pi}{2}},\hspace{0.5cm}<A^{2}>=\frac{59}{60},\hspace{0.5cm}
<A^{3}>=\frac{465}{512}\sqrt{\frac{\pi}{2}},....
\end{eqnarray}
 To get the results for the
small $\eta$ one needs to replace $A$ with $A-\frac{\eta}{2\pi}$ in
the (\ref{Area moments of Brownian meander}).

The next interesting case is the Neumann boundary with
$\eta\rightarrow\infty$. The Laplace transform of the distribution
of the area after a little algebra is
\begin{eqnarray}\label{PA meander N}
P^{\infty}(\lambda,T)=\sum_{i=1}^{\infty}\kappa_{i}e^{-2^{-1/3}\lambda^{2/3}c^{\infty}_{i}T},
\end{eqnarray}
where
$\kappa_{i}=\frac{AI(c^{\infty}_{i})}{c^{\infty}_{i}Ai(c^{\infty}_{i})}$
with $AI(z)=\int_{z}^{\infty}Ai(y)dy$. This distribution is exactly
the same as the distribution of the area of the Brownian motion,
i.e. $\int_{0}^{T}|B_{t}|dt$. This is not surprising because the
absolute value of Brownian motion is like considering the area in
the presence of totally reflecting boundary condition. Using the
inverse Laplace transform one can get
\begin{eqnarray}\label{PA meander N}
P^{\infty}(A,T)=\frac{2^{-1/3}T}{A^{5/3}}\sum_{i=1}^{\infty}\kappa_{i}c^{\infty}_{i}M_{\frac{2}{3}}(\frac{2^{-1/3}c^{\infty}_{i}T}{A^{2/3}})
\end{eqnarray}
Using the well known results for the area of the absolute value of
the Brownian motion \cite{Janson} the moments of the area can be
written as
\begin{eqnarray}\label{momentsPA meander N}
<A^{n}>=\frac{2^{-n/2}\Gamma(1+n)}{\Gamma(\frac{3n+2}{2})}L_{n},
\end{eqnarray}
where $L_{n}$ satisfies the following recursion relation
\begin{eqnarray}\label{recursion relation L}
L_{n}=\beta_{n}+\sum_{i=1}^{n}\frac{6j+1}{6j-1}\alpha_{j}L_{n-j},
\end{eqnarray}
and $\alpha_{j}$ and $\beta_{j}$ are the same as in the equation
(\ref{recursion relations Q}). The first few values are $L_{0}=1$,
$L_{1}=1$, $L_{2}=\frac{9}{4}$ and $L_{3}=\frac{263}{32}$. Then the
first few values of the moments of the area are
\begin{eqnarray}\label{Area moments of Brownian motion}
<A^{0}>=1,
\hspace{0.5cm}<A^{1}>=\frac{2}{3}\sqrt{\frac{2}{\pi}},\hspace{0.5cm}<A^{2}>=\frac{3}{8},\hspace{0.5cm}
<A^{3}>=\frac{263}{630}\sqrt{\frac{2}{\pi}},....
\end{eqnarray}

\subsection{Some applications}
\setcounter{equation}{0}\

Different applications of the area distributions of Brownian motion
in computer science, graph theory, fluctuating one-dimensional
interfaces and localization in electronic systems were already
discussed in length in many papers, see \cite{Majumdar,CDT} and
references therein. The Airy distribution function and its
derivative appear extensively in all applications. In this
subsection we will summarize some immediate simple applications of
our extended Airy distribution. We will discuss the distribution of
the average distance of a  particle from a disorder with point
interaction in three dimensions and the distribution of the average
relative height distribution of interacting interfaces in two
dimensions. We will also discuss one more elastic Brownian
functional distribution related to vicious walkers interacting with
the boundary.

The first immediate application comes from the equality of the norm
of the ~3-dimensional Brownian motion (called three dimensional
Bessel process) and Brownian motion on the half line
\cite{RY,Rajabpour}. This is very easy to show by considering the
radial part of the Fokker-Planck equation of three dimensional
Brownian motion. Now consider a Brownian motion moving in three
dimensions in the presence of the disorder at the origin ,  one can
map the system to the problem of one particle on the half line. The
most generic point interaction between disorder and  particle comes
from the self adjoint extension of the Hamiltonian of the free
particle in punctured three dimensional space \cite{Rajabpour} which
is equal to the free particle on the half line with the boundary
condition that we discussed in section ~2. Then it is easy to see
that the area distribution that we calculated is just the average
distance distribution in the period $T$ between the disorder and the
free particle with the generic point interaction.

Another simple application, which is in the close connection with
the previous example, is the interacting interfaces. A path of
Brownian motion in the $x-t$ space is just like an interface with
zero roughness exponent. One can also look at this interface as an
ensemble of growth models such as Edwards-Wilkinson model. Consider
now two non-crossing interfaces in the region $[0,L]$ with the
similar boundary conditions. This problem is equivalent to the
problem of two free particles on the real line. Consider, like the
previous example, point interactions between the particles. The
interaction between the particles is equivalent to the interaction
between the interfaces. Then the area distribution that we
calculated for the elastic Brownian excursion is just the
distribution of the average distance of the interfaces in the
interval $[0,L]$. One can also relax the boundary condition in one
of end points and consider different boundary conditions for the
different interfaces and use the results corresponding to elastic
Brownian meander area.

Since the elastic Brownian motion is the generalization of Brownian
motion in the presence of the boundary we believe that all the
possible applications should deal with the boundary interaction or
point interaction between two particles. In this paper we just
calculated one possible functional of the elastic Brownian motion,
the area. However, there are many other functionals that can have
interesting physical applications in the study of the interacting
non-crossing walkers or interacting interfaces. We will discuss some
of these functionals in the conclusion of the paper and give here
one more example with more detail.

Consider the problem of $p$ non-crossing walkers, which has
application in  describing domain walls of elementary topological
excitations in the commensurate adsorbed phases close to the
commensurate-incommensurate transition \cite{HF}, in the presence of
the boundary. One interesting quantity is the maximal height
distribution of the top walker that was already calculated exactly
for the vicious walkers in \cite{SMCF}. Walkers are vicious if they
annihilate each other when they meet. For simplicity we will discuss
the simplest case $p=1$. Consider $H$ as the maximal height of the
walker in $[0,1]$ then one can define the cumulative distribution as
$P(M)=Prob[H\leq M]$. We define $N(\epsilon,M)$ as the probability
that the  walker do an excursion in the period $[0,1]$ starting at
$\epsilon$  and coming back to the same point staying within the
interval $[0,M]$. The cut-off $\epsilon$ is just necessary for the
Dirichlet boundary condition as we discussed before. Since for this
case the result is already known \cite{chung} we will put
$\epsilon=0$ hereafter. Then it is easy to show that
$N(0,M)=\sum_{E}|\psi_{E}|^{2}e^{-E}$. The wave function is as
(\ref{Wave function momentu}) , i.e.
$\psi_{E}=\sqrt{\frac{2}{\pi}}\cos(kx+\delta_{k})$, with
$E=\frac{k^{2}}{2}$ and $k$ is the solution of the equation $\cot
(kM)=\frac{2\pi}{\eta k}$. Finally one can write the cumulative
distribution as
\begin{eqnarray}\label{cumulative}
P(M)=\frac{1}{Z^{\eta}}\sum_{k}\frac{1}{(1+(\frac{2\pi}{\eta
k})^{2})(\frac{M}{2}-\frac{1}{4k}\sin(2kM))}e^{-k^{2}/2},
\end{eqnarray}
where $Z^{\eta}$ comes from the equation (\ref{ZE2}). For small
$\eta$ and small $M$ with $M <\eta$ one can simplify the equation as
$P(M)\approx \frac{\sqrt{2\pi}}{M}k_{0}^{2}e^{-k_{0}^{2}/2}$, where
$k_{0}=\frac{\sqrt{3-\frac{6\pi M}{\eta}}}{M}$.

Generalization of the above results to arbitrary $p$ is
straightforward, however, it is rather cumbersome.
 From the perspective of our study in this paper one can
generalize this problem in two directions:  the first possibility is
considering non-vicious walkers, interacting domain-walls, with a
Dirichlet boundary condition on the wall. The second possibility is
considering vicious walkers with non-trivial interaction with the
boundary.

\section{The area of the Brownian bridge on the line with a point defect }\
\setcounter{equation}{0}\

In this section we will summarize some results come from the area
calculation for the Brownian bridge on the line with a point defect.
The quantum mechanical counterpart was discussed lengthly in the
literature and it is called the free particle on a line with a hole
\cite{Albeverio}. The functional integral for this problem was
discussed in \cite{gutman2} and it is based on the different local
times of the particle in the two sides of the origin. The most
general boundary condition that respects the time reversal symmetry
for the defect on the origin, after using self adjoint extension
theory, is
\begin{eqnarray}\label{bc}
\left(
  \begin{array}{c}
    \psi'_{+}(0) \\
    \psi_{+}(0)
  \end{array}
  \right) =
  \begin{pmatrix}
  -\alpha & -\beta  \\
  -\delta & -\gamma  \\
   \end{pmatrix}\left(
  \begin{array}{c}
    \psi'_{-}(0) \\
    \psi_{-}(0)
  \end{array}
  \right),
\end{eqnarray}
with a constraint $\alpha\gamma-\beta\delta=1$. For simplicity we
will consider some special cases. It is easy to see that for
$\delta\rightarrow\infty$ the boundary condition decouples and we
will have two decoupled half lines and so the area distribution is
as the previous section.

The next simple case is $\alpha=\gamma=-1$ and $\delta=0$ which is
equal to the free particle on the line with the following delta
function potential
\begin{eqnarray}\label{delat function potensial}
V(x)=-\frac{\beta}{2}\delta(x).
\end{eqnarray}
To avoid the bound state we consider non-positive $\beta$. To
calculate the area of this kind of Brownian bridge one can use the
method of previous section. Interestingly the results are very
similar to the previous section.  The energy of the particle is
$E_{k}=\frac{1}{2}k^{2}$ and the wave functions are
\begin{eqnarray}\label{Wave function momentum delta}
\psi_{k}(x)=\sqrt{\frac{1}{\pi}}\cos(k|x|+\delta_{k}),
\end{eqnarray}
where $\tan(\delta_{k})=\frac{\beta}{2 k}$. Then $Z^{\beta}_{E}$ is
the same as (\ref{ZE2}) after replacing $\eta$ with
$\frac{4\pi}{\beta}$. The wave functions in the presence of the
potential $\lambda |x|$ can have different parities, they are
\begin{eqnarray}\label{Wave function airy and delta}
\psi^{\beta}_{i}(x)&=&\frac{2^{-1/2}(2\lambda)^{1/6}Ai((2\lambda)^{1/3}|x|-c_{\beta i})}{\sqrt{c_{\beta i}Ai^{2}(-c_{\beta i})+Ai'^{2}(-c_{\beta i})}},\hspace {2cm}\hbox{even parity},\\
\psi_{i}(x)&=&
\hbox{sgn}(x)2^{-1/2}(2\lambda)^{1/6}\frac{Ai((2\lambda)^{1/3}|x|-c_{-\infty
i})}{Ai'(-c_{-\infty i})},\hspace {0.5cm}\hbox{odd parity},
\end{eqnarray}
where $c_{\beta}$ is the same as $c^{\eta}$ after replacing $\eta$
with $\frac{4\pi}{\beta}$. Odd wave functions do not contribute in
the distribution of the area. The result for the even part is the
same as the result for the elastic Brownian excursion. It is easy to
see that $\beta=0$ is like the free Brownian motion and so the
distribution of the area is like Brownian bridge or like elastic
Brownian excursion with Neumann boundary condition, i.e. (\ref{Pa
Neumann}). The $\beta=-\infty$ is like the the Dirichlet boundary
condition and the distribution of the area is (\ref{alfa2 pa}) after
considering $\eta=0$.

The other simple boundary condition comes from $\alpha=\gamma=-1$ ,
$\beta=0$ and $\delta\neq 0$, in the other words
\begin{eqnarray}\label{bc non-trivial}
\psi'_{+}(0)-\psi'_{-}(0)&=&0,\\
\psi_{+}(0)-\psi_{-}(0)&=&-\delta\psi'_{-}(0).
\end{eqnarray}
The energy of the particle is $E_{k}=\frac{1}{2}k^{2}$ and the wave
functions are
\begin{eqnarray}\label{Wave function momentum delta}
\psi_{k}(x)=\hbox{sgn}(x)\sqrt{\frac{1}{\pi}}\cos(k|x|+\delta_{k}),
\end{eqnarray}
where $\tan(\delta_{k})=\frac{2}{\delta k}$. Then $Z^{\delta}_{E}$
is the same as (\ref{ZE2}) after replacing $\eta$ with $\pi\delta$.

The wave functions in the presence of the potential $\lambda |x|$
can have different parities, they are
\begin{eqnarray}\label{Wave function airy and delta}
\psi^{\delta}_{i}(x)&=&\hbox{sgn}(x)\frac{2^{-1/2}(2\lambda)^{1/6}Ai((2\lambda)^{1/3}|x|-c^{\pi\delta}_{i})}{\sqrt{c^{\pi\delta}_{
i}Ai^{2}(-c^{\pi\delta}_{ i})+Ai'^{2}(-c^{\pi\delta}_{ i})}},\hspace
{0.5cm}\hbox{odd parity},\\
\psi_{i}(x)&=&
(2\lambda)^{1/6}\frac{Ai((2\lambda)^{1/3}|x|-c^{\infty}_{
i})}{\sqrt{2c^{\infty}_{i}}Ai(-c^{\infty}_{i})},\hspace
{2cm}\hbox{even parity}.
\end{eqnarray}
The above calculation shows that in this case the distribution of
the area can be calculated by adding two terms. In this case the
even parity has contribution for the distribution of the area which
is equal to the Neumann boundary condition. Replacing $\eta$ with
$\pi\delta$ in the formula of elastic Brownian excursion will give
the contribution of the odd part. The case of
$\delta\rightarrow\infty$ is equal to the Neumann boundary condition
that separates the system in to the two regions, positive and
negative part of the real line. Of course having two solutions is an
indicator of degeneracy coming from the parity invariancy of the
system in this limit.  $\delta=0$ is the free particle case and one
can see that only the even parity has contribution in the area
distribution.

\section{Conclusion and discussion }\
\setcounter{equation}{0}\

In this paper we found the area distribution of the elastic Brownian
motion in some limits. Our method was based on the equality of this
process with the self adjoint Hamiltonian of the quantum particle on
the half line. The corresponding Hamiltonian for the area
distribution is just the Hamiltonian with linear potential. The
eigenvalues of this Hamiltonian after self adjoint extension satisfy
a transcendental equation and so for the generic case the
distribution of the area is not available, however, in some limits
the calculation is tractable. We found perturbatively the area
distribution of the Brownian excursion and the Brownian meander in
the presence of the weekly reflecting barrier. By using self adjoint
extension we found a unified method to classify different possible
area distribution for the Brownian motion in the presence of the
boundary. Some possible applications in diffusion phenomena in the
presence of disorder and interacting interfaces were also discussed.

We did similar calculations for the Brownian motion on the pointed
line. The self adjoint Hamiltonian in this case has three
independent parameters and the eigenvalues of the Hamiltonian
satisfy the same transcendental equation in some interesting limits.
Similar calculations can be useful in describing different
distributions of diffusing particles in the presence of point
disorder.

There are plenty of questions remained to be answered in the study
of the elastic Brownian motion by using quantum mechanical technics.
Some of them are:  The case of the maximal height of $p$
non-intersecting Brownian excursions and Brownian bridges is also
interesting to be calculated \cite{SMCF}, the possible connection of
this study to the interacting domain walls of elementary topological
excitations in the commensurate adsorbed phases close to the
commensurate-incommensurate transition can be interesting. The other
example is the distribution of the time to reach the maximum
\cite{MFKY}. Unfortunately the eigenvalue equations for the above
cases are transcendental as we faced for one example in the end of
section ~3 and so it is impossible to get a closed formula for the
distributions, however, the exact calculations in some limits are
may be possible. The other example is the distribution of the time
spent by the particle on the positive side of the origin out of the
total time $t$. This distribution was first calculated by Levy in the
case of the Brownian motion \cite{Levy}. For the pointed line the
equations are again transcendental and need to be solved by
numerical calculation. We mostly focused on the distributions in one
dimension however one can also try to calculate the different
distributions like the algebraic area distribution, winding number
distribution in the pointed two dimensional space, the number of
defects could be finite or infinite \footnote{For one defect point
we don't expect significant change in the winding number
distribution in two dimensions.}. We believe that the method of self
adjoint extension in quantum mechanics can be useful to calculate
such kind of distributions. It is also interesting to check our
results with the numerical calculations.

\end{document}